\documentstyle[11pt,newpasp,twoside,psfig]{article}
\def\eq#1 {\begin{equation} #1 \end{equation}}

\def\non  {\nonumber \\}
\def\x    {\hbox{$\times$}}
\def\about{\hbox{$\sim$}}
\def\sub#1{_{\rm #1}}
\def\vk   {\hbox{${\bf v}\sub{k}$}}
\def\Js   {\hbox{$J\sub{s}$}}
\def\k    {\hbox{$\kappa$}}
\def\kl   {\hbox{$\kappa_l$}}
\def\kc   {\hbox{$\kappa_c$}}
\let \half=\onehalf
\let \third=\onethird
\let \quarter=\onequarter
\def\nuB      {\hbox{$\nu_B$}}
\def\DnuD     {\hbox{$\Delta\nu_D$}}
\def\PI       {\hbox{${\bf \Pi}$}}
\def\kpvec    {\mbox{\boldmath $\kappa_p$}}
\def\xs       {\hbox{$x_s$}}
           \let\xb=\xB
\def\Dm       {\hbox{$\Delta m$}}
\def\Ob       {\hbox{$\Omega_{b}$}}
\def\Js       {\hbox{$J_s$}}
\def\Epar     {\hbox{$E_\|$}}

\def\Tpar     {\hbox{$T_\|$}}
\def\Tperp    {\hbox{$T_\bot$}}
\def\H2O{\hbox{H$_2$O}}

\begin{document}
\title{Topics in Basic Maser Theory}
 \author{Moshe Elitzur}
\affil{Physics \& Astronomy Department, University of Kentucky, Lexington,
       KY 40506, USA; moshe@pa.uky.edu}

\begin{abstract}
This review covers some of the developments in basic theory of astronomical
masers over the past ten years. Topics included are the effects of three
dimensional geometry and polarization, with special emphasis on the
differences between maser and non-maser radiation.
\end{abstract}

\section{Maser Absorption Coefficient in Three Dimensions}

The absorption coefficient \k\ describes the coupling between particles and
radiation and serves as the foundation of maser theory. The appropriate
expression, taking proper care of the Doppler matching between particles with
thermal velocity distribution and photons with a given wave vector {\bf k}
(aligned with the $z$-axis) and frequency $\nu$, was developed by Litvak
(1973). It can be written as
\eq{\label{eq:Litvak}
 \k({\bf k}) = {c\over\nu_0}\int\!\!{\k_0(\vk)\over 1 + I(\vk)/\Js}\ dv_x dv_y\,,
}
where
\[
    \vk = (v_x, v_y, v_z = c[\nu - \nu_0]/\nu_0), \qquad
    I(\vk) = {1\over4\pi} \int I(\bf k')\delta(\nu' - \nu_0 + \vk\cdot{\bf k'})
                            d^3{\bf k'}.
\]
Here $I(\bf k)$ is the intensity for wave vector {\bf k}, \Js\ is the
saturation intensity and $\k_0(\bf v)$ is the unsaturated absorption
coefficient (proportional to the density of particles with thermal
velocity~{\bf v}). Unfortunately, this expression was soon ignored and largely
forgotten (except for a couple of papers by Bettwieser \& Kegel 1974 and
Bettwieser 1976) because shortly thereafter Goldreich \& Kwan (1974) introduced
the much simpler expression
\eq{\label{eq:Standard}
    \k_\nu = {\k_{0\nu}\over1 + J_\nu/\Js}\,,
}
where
\[
    \k_{0\nu} = {c\over\nu_0}\int\!\!\k_0(\vk)\ dv_x dv_y\,, \qquad
    J_\nu = \int I_\nu{d\Omega\over4\pi}\,.
\]
This expression gives the correct result for linear masers, where photon and
particle motions are aligned, and because maser radiation is tightly beamed,
Goldreich \& Kwan reasoned that it should be adequate for all masers.   It
became the standard for all subsequent theory, including 3D geometries, even
though it was not derived from the proper equation \ref{eq:Litvak}. It took
many years until Neufeld (1992) recognized the internal inconsistency in this
approach: in deriving the standard expression (\ref{eq:Standard}) the maser
beaming angle is assumed to vanish, yet this very expression is used to solve
the maser structure in any given geometry and derive the beaming angle for that
geometry.

Fortunately, the standard expression proved to be mostly adequate. Since maser
radiation is tightly beamed, equation \ref{eq:Litvak} can be handled with the
aid of a series expansion in the beam width and the leading term in such a
series reproduces the standard expression 2 (Elitzur 1994). The standard theory
provides the correct description of three dimensional masers and its results
remain intact but only within the frequency core $|x| \la \xs\DnuD$, where
\DnuD\ is the Doppler width,  $x = (\nu - \nu_0)/\DnuD$ and \xs\ is a
dimensionless parameter.  For typical pumping schemes \xs\ is \about\ 2 in
spherical masers, \about\ 2.5--3 in disk masers and \about\ 3--5 in cylindrical
masers.  For frequencies outside this core region, interaction with core rays
that are slightly slanted to the direction of propagation suppresses photon
production. Observed maser radiation is effectively confined to the core
region, frequencies in the suppressed domain are essentially unobservable.  In
practice, suppression only affects extreme maser outbursts. Their profiles
change in such a way that when fitted with a Gaussian they mimic line narrowing
in proportion to $(\ln F_0)^{-1/2}$, where $F_0$ is the flux at line center, in
contrast with the standard theory where such behavior is confined to
unsaturated amplification.  Such an inverse correlation between intensity and
linewidth has been detected in a number of \H2O maser flares in star-forming
regions (e.g.\ Mattila et al 1985, Rowland and Cohen 1986, Boboltz et al 1993,
Liljestrom 1993).

\subsection{Linewidths in 3D Masers}

Maser radiation is tightly beamed, therefore the intensity $I_\nu$,
angle-averaged intensity $J_\nu$ and flux $F_\nu\ (= \int \mu I_\nu d\Omega)$
are related via
\eq{
    F_\nu = 4\pi J_\nu = I_\nu\Ob
}
where \Ob\ is the beaming solid angle. Saturated masers display two types of
beaming (Elitzur, Hollenbach \& McKee 1992). In matter-bounded masers, whose
prototype is the filamentary maser, the beaming angle is controlled by the
matter distribution and the maser observed size is equal to its physical size.
The beaming angles of such masers are frequency independent, therefore the
frequency profiles are the same for $F_\nu$ and $I_\nu$. In
amplification-bounded masers, whose prototype is the spherical maser, the
beaming angle is controlled by the amplification process and the observed size
is significantly smaller than the physical size. Because the amplification is
strongest at line center the beaming is tightest there; the beaming angle
increases with frequency shift from line center and the spectral shapes of
$F_\nu$ and $I_\nu$ are different from each other.

With the standard expression for the absorption coefficient
(\ref{eq:Standard}), it is easy to show that the flux of a saturated maser
increases with length $\ell$ according to $F_\nu \propto \k_{0\nu}\Js\ell$
independent of the geometry (Elitzur 1990). Even with the full, proper equation
\ref{eq:Litvak}, the standard theory remains applicable at the core of the line
so this result too is valid there. Therefore, in any saturated maser the
frequency profile of the flux always obeys $F_\nu \propto \k_{0\nu} \propto
\exp(-x^2)$, i.e., the flux spectral shape displays the full Doppler width
\DnuD. On the other hand, the intensity obeys $I_\nu \propto F_\nu/\Ob$,
therefore its spectral pprofile will reflect also the frequency dependence of
\Ob. The result is $I_\nu \propto \k_{0\nu}^\alpha \propto \exp(-\alpha x^2)$
where $\alpha$ is the dimensionality of the geometry; $\alpha$ is 1 for linear
or filamentary masers, 2 for planar masers such as disks and 3 for fully 3D
structures such as spheres. The width of the brightness spectral shape is thus
\eq{
    \Delta\nu = {\DnuD\over\sqrt{\alpha}}.
}
In particular, the intensity linewidth of a saturated planar maser (the most
likely geometry for shock induced masers) is 40\% smaller than the Doppler
width.

\section{Polarization}

Thermal radiation is generated in spontaneous decays, maser radiation in
stimulated emission.  This fundamental difference has profound
implications, especially for polarization.

The stimulated emission process is the inverse of radiation absorption.
Absorption is a purely classical process, therefore the same applies also to
stimulated emission. Since line radiation involves discrete energy states, the
particle properties must be described with quantum mechanics. But the radiation
wavelength is many orders of magnitude larger than particle dimensions so there
is no need to quantize also the radiation field. The interaction of matter with
maser radiation is adequately described with a hybrid, semi-classical approach
(Litvak 1970; Goldreich, Keeley, \& Kwan 1973, GKK hereafter): The radiation
field is described by standard classical electromagnetic waves. The energy
levels are eigenstates of the system Hamiltonian, treated by quantum theory.
Interaction with the radiation field, treated as a perturbation, causes
transitions between the energy levels. The transition rates for both absorption
and stimulated emission are obtained from the product of the (quantum) matrix
element of the transition dipole moment with the (classical) intensity of the
radiation field. Since this is a complete description, not merely a classical
analog, an important consequence is that there are no properties of the
radiation generated in stimulated emission that are peculiar to the quantum
theory; we must be able to fully deduce all of them from purely classical
concepts as applied to propagating electromagnetic waves.

In contrast, spontaneous emission is a purely quantum process. It has no
classical analog since the initial state is devoid of radiation and thus cannot
interact with any electromagnetic wave. Spontaneous decays do not occur
even in standard treatments of quantum theory because the energy levels are
stationary states of the system Hamiltonian, completely stable in the absence
of external perturbations. This process occurs only when quantization of the
electromagnetic field is taken into considerations, and can be interpreted as
scattering off vacuum fluctuations. Spontaneous emission can be analyzed only
in terms of the photon description of the radiation field.

\paragraph{Induced Photons} When stimulated emission is described in terms of
photons, energy and momentum conservation imply that the induced photon has the
same frequency and direction as the parent photon. However, contrary to some
widespread misconceptions, {\em the induced photon does not have the same}

\begin{enumerate}
\item {\bf Phase:} The argument of the oscillatory behavior of any wave is its
phase $\phi = \phi_0 + \omega t -\, {\bf k\!\cdot\!r}$, where $\omega$ is the
angular frequency and {\bf k} is the wave vector. An electromagnetic wave has a
phase, a photon does not. The uncertainty principle leads to the relation
$\Delta n\Delta\phi \geq 1$ between phase and photon number. The phase of a
state with a well-defined number of photons is completely undetermined. {\em
Phases are meaningless when dealing with photon numbers}, in particular they
are irrelevant in spontaneous emission.

\item {\bf Polarization:} The induced photon polarization is not necessarily
equal to that of the parent photon. Instead, it is determined by the change in
magnetic quantum number $m$ of the interacting particle. \Dm\ = 0 transitions
couple to photons linearly polarized along the quantization axis, $\Dm = \pm1$
transitions couple to photons that are right- and left-circularly polarized in
the plane perpendicular to the quantization axis. Consider the interaction of a
linearly polarized photon with particles in the upper level of a spin $1 \to 0$
transition. When the interacting particle is in the $m = 0$ state it executes a
\Dm\ = 0 transition and the induced photon, too, is linearly polarized. But
this is not the case when the particle is in one of the $|m|$ = 1 states. The
linearly polarized photon, which can also be described as a coherent mixture of
two circularly polarized photons, will now induce a $|\Dm| = 1$ transition and
the induced photon is circularly polarized. {\em Induced emission preserves
polarization only when the magnetic transitions do not overlap}.
\end{enumerate}

\subsection{Polarization in Spontaneous Decays}

\begin{figure}
\centering \leavevmode
 \psfig{file=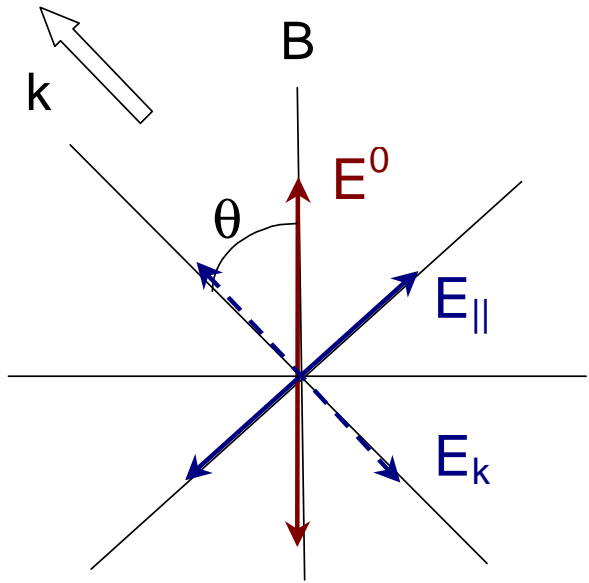,width=0.6\hsize,clip=}
\caption{Polarization in \Dm\ = 0 spontaneous emission.}
\label{fig:Dm0Polarization}
\end{figure}

From Maxwell's equations, the electric field of an electromagnetic wave is
always perpendicular to the propagation direction.  This seemingly simple
transverse condition is a rather peculiar constraint, hard to reconcile with
the properties of the quantized particles that emit line radiation. Figure
\ref{fig:Dm0Polarization} shows the geometry for a \Dm\ = 0 spontaneous decay.
The quantization axis is denoted by $B$. The electric field generated in the
transition, with an amplitude $E^0$, is always parallel to this axis. The
photons propagate in the direction marked by the double arrow at an angle
$\theta$ from $B$, the corresponding axis is denoted $k$. The component of the
electric field along the axis parallel to the projection of $B$ on the plane of
the sky is $\Epar = E^0\sin\theta$. What about the component along the
direction of propagation, $E_k$? Is it $E^0\cos\theta$ as the geometry
dictates? Or is it 0 as required by the transverse condition? How can we
reconcile these conflicting results? The answer is that we cannot as long as we
apply classical reasoning. The resolution of this conflict is rooted in the
quantum nature of spontaneous emission, which has no classical analog. Because
of the uncertainty principle, only one component of any vector can be
determined whenever the magnitude of that vector is known, the other two remain
undetermined; recall the properties of angular momentum. The longitudinal
component can be ignored in spontaneous emission---quantum mechanics can be
counted on to take care of the transverse condition $E_k = 0$ and we can
proceed directly to calculate the polarization. Denote by $I^0$ the intensity
associated with the amplitude $E^0$ ($I^0 \propto |E^0|^2$). Then the
intensities measured by a linear antenna oriented parallel and perpendicular to
the $B$-axis are, respectively, $\Tpar = I^0\sin^2\theta$ and $\Tperp = 0$.
Therefore, the Stokes parameters of \Dm\ = 0 spontaneous emission are $I =
\Tpar + \Tperp = I^0\sin^2\theta$ and $Q = \Tpar - \Tperp = I$, recovering the
standard result of full linear polarization.

\subsection{Fully Resolved Zeeman Pattern; $\nuB \gg \DnuD$}
\label{sec:xb > 1}

When the magnetic field is sufficiently strong that the Zeeman shift \nuB\
exceeds the linewidth \DnuD, radiation is produced in pure \Dm\ transitions
centered on the appropriate Zeeman frequencies. For the \Dm\ = 0 transition we
have just derived the polarization and it is straightforward to repeat these
calculations for $\Dm = \pm1$ spontaneous emission. The results are summarized
in table \ref{table:Zeeman polarization} for the classical Zeeman pattern.  In
that case there are three spectral lines centered on $\nu_0 + \nuB\Dm$ (\Dm\ =
0, $\pm1$), where $\nu_0$ is the line frequency in the absence of a magnetic
field, with $I^0(\nu) = I^+(\nu + \nuB) = I^-(\nu - \nuB)$. Quantities listed
in the first column are obtained for each transition from the product of the
intensity heading the transition column with the appropriate trigonometric
factor. The intensities that would be measured with right- and left-circular
instrumental response are listed as $T_{r,\,l}$ and $V = T_r - T_l$. The
parameter $U$ vanishes for all transitions with this choice of axes.

\begin{table}
  \begin{center}
    \begin{tabular}{l|ccc}
                        & $\sigma^+$ & $\pi$    & $\sigma^-$              \\
                        \\
                        & $\quarter I^+$
                        & $\half    I^0$
                        & $\quarter I^-$                     \\
                        \tableline
\\
    $\Tpar$ & $2\cos^2\theta$    & $2\sin^2\theta$ & $2\cos^2\theta$      \\
    $\Tperp$& 2                  & 0               & 2                    \\
    $T_r$   & $(1+\cos\theta)^2$ &  $\sin^2\theta$ & $(1-\cos\theta)^2$   \\
    $T_l$   & $(1-\cos\theta)^2$ &  $\sin^2\theta$ & $(1+\cos\theta)^2$   \\
\\
    $I$     & $2(1+\cos^2\theta)$& $2\sin^2\theta$ & $2(1+\cos^2\theta)$  \\
    $Q$     & $-2\sin^2\theta$   & $2\sin^2\theta$ & $-2\sin^2\theta$     \\
    $V$     & $4\cos\theta$      & 0               & $-4\cos\theta$       \\
    \hline
    \end{tabular}
  \end{center}
\caption{Polarizations for fully resolved Zeeman pattern, $\nu_B \gg \DnuD$}
\label{table:Zeeman polarization}
\end{table}

The results display the standard polarization properties of thermal radiation
of fully resolved Zeeman components. And because each component can be
considered an independent, isolated radiative transition that couples to a
single sense of polarization, these results apply also to maser radiation even
though they were derived for spontaneous emission. Indeed, these are the maser
polarization properties derived by GKK, although from an entirely different
approach. {\em Thermal and maser polarizations are the same when the Zeeman
pattern is fully resolved}. The only difference between the two cases is the
disparity between the $\pi$ and $\sigma$ maser intensities, reflecting their
different growth rates (Elitzur 1996). This disparity predicts a preponderance
of $\sigma$-components.  However, the evidence is mounting that individual
$\pi$-components are in fact never observed, a puzzle that currently has no
explanation.

\subsection{Overlapping Zeeman Components; $\nuB \ll \DnuD$}

The thermal and maser cases diverge now because the stimulated emission mixes
the polarization components. Thermal emission is produced in spontaneous decays
and must be considered in the photon picture; only intensities, i.e., photon
numbers, are relevant. Maser radiation is generated in stimulated emission and
its properties must be understood in terms of classical waves interacting with
particles in quantized energy levels; both amplitudes and phases count.

\paragraph{Thermal Radiation}
The various \Dm\ transitions produce spectral components with equal intensities
but centered on frequencies slightly shifted from each other. The intensity
$I^0$ of the \Dm\ = 0 component is an even function of $x$, centered on $x =
0$. Introduce $\xb = \nu_B/\DnuD \ll 1$,
then $I^\pm(x \pm \xb) = I^0(x)$. The three components are produced independent
of each other and the overall radiation field is an incoherent superposition of
them. Within each component the transverse condition is obeyed independently
and the photons are polarized as described in table \ref{table:Zeeman
polarization}. Adding up the Stokes parameters at every frequency across the
line yields
\begin{eqnarray}\label{eq:Zeeman}
   I &= &I^0 + I^+ + I^- = 2I^0 \non
   Q &= &Q^0 + Q^+ + Q^- = \left[I^0 - \half(I^+ + I^-)\right]\sin^2\theta
                          = -{d^2I\over dx^2}\,(\xb\sin\theta)^2 \non
   V &= &V^0 + V^+ + V^- = (I^+ - I^-)\cos\theta
                          = {dI\over dx}\,\xb\cos\theta
\end{eqnarray}
The final expression in each case is the leading order result from a series
expansion in \xb\ of $I^\pm(x)$ around $I^0(x)$. The overall radiation field is
polarized because it is the superposition of three polarized {\em different}
intensities---because of the Zeeman shifts the intensities are slightly
different at every given frequency.  These differences are controlled by the
parameter \xb, and so is the polarization. When $\xb\ \to 0$, the polarization
disappears.

\paragraph{Overlapping Electromagnetic Waves} \label{sec:xb < 1}

\begin{figure}
\centering \leavevmode
 \psfig{file=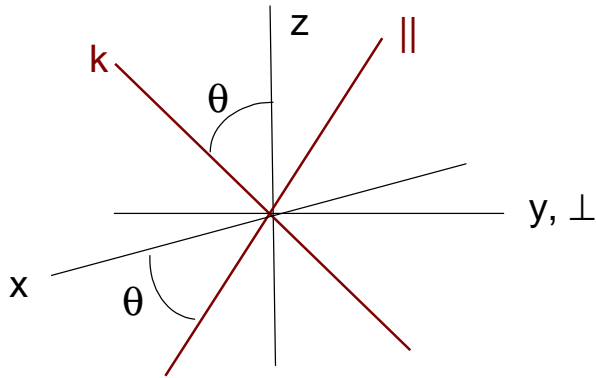,width=0.6\hsize,clip=}
\caption{Geometry of the wave and quantization frames.} \label{fig:B-k}
\end{figure}

At a given frequency and wave vector, each \Dm\ (= $0, \pm1$) transition
produces an electric vector with magnitude
\eq{\label{eq:components}
    E^{\Delta m} = |E^{\Delta m}|\,
                   e^{i({\bf k\cdot r} - \omega t + \phi_{\Delta m})},
}
where the initial phase $\phi_{\Delta m}$ is random. The overall electric
vector is ${\bf E} = \sum{{\bf E}^{\Delta m}}$ and must obey the transverse
condition ${\bf E\cdot k} = 0$. The geometrical setup is shown in figure
\ref{fig:B-k}. Particle quantization is defined with respect to the $x$-$y$-$z$
coordinate frame, with $z$ the quantization axis. In this frame, each
component of {\bf E} is uniquely associated with a specific \Dm: the
$z$-component couples only to \Dm\ = 0, so that $E^0 = E_z$, the $x$- and
$y$-components couple only to $\Dm = \pm1$, i.e., $E^{\pm} = 2^{-1/2}(E_x \pm
iE_y)$. The wave propagation is along the $k$-axis, rotated by an angle
$\theta$ from the quantization axis in the $x$-$z$ plane. Since the electric
field is a proper vector, it can be decomposed in this frame too using
straightforward, standard geometry. The transverse condition states that $E_k =
0$ irrespective of the direction of propagation, and now we cannot rely on
quantum considerations; this condition must be obeyed as a geometrical
constraint on the three vector components of {\bf E} (eq.\
\ref{eq:components}). Given the amplitudes $|E^{\Delta m}|$, the condition $E_k
= 0$ becomes a relation among the phases $\phi_{\Delta m}$. For example, when
$|E^0| = |E^+| = |E^-|$ the relation is
\begin{figure}
\centering \leavevmode
 \psfig{file=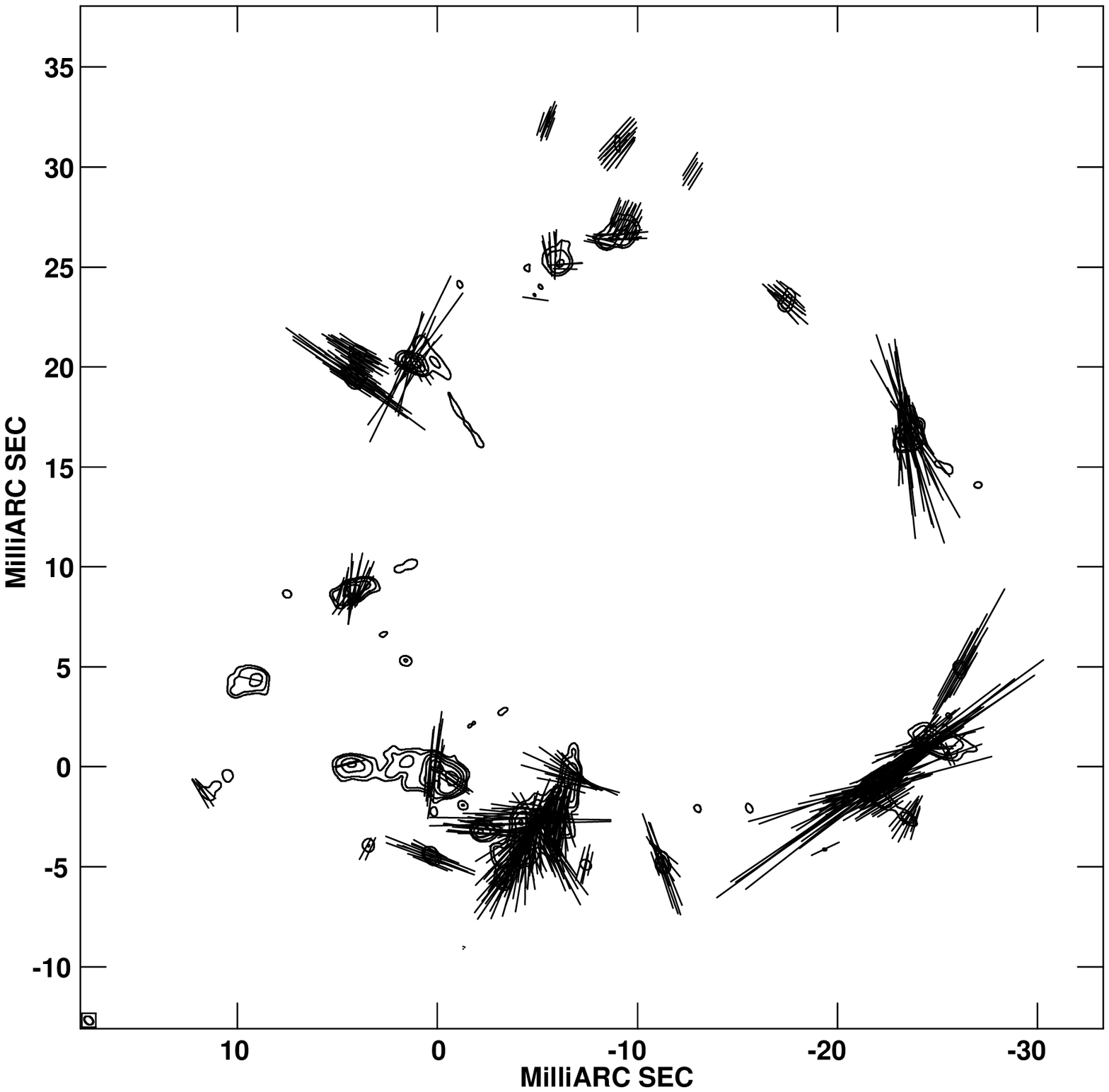,width=0.4\hsize,%
  bbllx=170pt,bblly=260pt,bburx=265pt,bbury=365pt,angle=270,clip=}
\caption{90\deg\ flip of SiO linear polarization in TX Cam (Kemball \& Diamond
1997)} \label{fig:flip}
\end{figure}
\eq{
        \phi_+ = -\phi_-\,,         \qquad
        \phi \equiv |\phi_\pm| =  \arccos\left(2^{-1/2}\cot\theta\right),
}
where the meaningless overall phase is set through $\phi_0 = 0$ (Elitzur 1991);
one $\sigma$-component leads the $\pi$-component by the phase difference
$\phi$ and the other must trail by the exact same amount. Only waves launched
with these phase relations produce superpositions that are purely transverse so
that they can be amplified by propagation in the inverted medium. The
transverse components of these propagating waves are linearly polarized
according to
\eq{\label{eq:GKK}
  q = {Q\over I} = -1 + {2\over3\sin^2\theta}\ .
}
Whereas the thermal polarization arises from the superposition of different
intensities, this one involves equal intensities but well defined phase
relations among the amplitudes. The polarization arises because the independent
constraints imposed by the particle interactions and the transverse condition
must be reconciled simultaneously. Particle interactions (i.e., maser pumping)
produce three independent fields corresponding to \Dm\ = $0, \pm1$. The
transverse condition dictates that only two independent fields propagate in any
given direction, the longitudinal combination of the original fields must
vanish. The resulting phase relation, and polarization, reflect the correlation
that must exist to eliminate the longitudinal component of {\bf E}. The linear
polarization in eq.\ \ref{eq:GKK} depends only on propagation angle, it is
entirely independent of \xb. Indeed, the only assumption in its derivation was
the existence of a quantization axis in the source---the physical process
behind this axis was never specified, in principle it need not be a magnetic
field.

Equation \ref{eq:GKK} is immediately recognized as the polarization solution
derived by GKK from an entirely different approach. Their assumption of equal
pump rates for the different $m$-states is reflected in the equal $|E^{\Delta
m}|$ taken here. The polarization becomes unphysical ($q > 1$) for
$\sin^2\theta < \third$. Only unpolarized maser radiation can propagate there
since the interference dictated by the transverse condition cannot be obeyed
for equal $|E^{\Delta m}|$. The polarization changes sign at $\sin^2\theta =
\twothirds$, where it vanishes. At smaller (larger) angles $q$ is positive
(negative), corresponding to polarization along (perpendicular to) the
quantization axis. The transition between positive and negative $q$ corresponds
to a 90\deg\ flip in the polarization direction. Such flips are commonly
observed in SiO masers; an example is shown in figure \ref{fig:flip}. A natural
explanation is a slight change in direction of the magnetic field, straddling
the two sides of the transition angle $\theta = 55\deg$.

It is important to note that the linear polarization will usually not exceed
33\% because only the limited range $35\deg \le \theta \le 45\deg$ gives $|q| >
\third$. Propagation at $\theta > 45\deg$ gives only $|q| \le \third$---along
the field when $45\deg \le \theta < 55\deg$ and orthogonal to it when $\theta >
55\deg$. The direction of the magnetic axis is generally not known a-priori. It
can be determined with certainty only when the linear polarization exceeds
33\%, in which case the field projection on the plane of the sky must be
parallel to the polarization.

\paragraph{Maser Polarization}

\begin{figure}
\centering \leavevmode
 \psfig{file=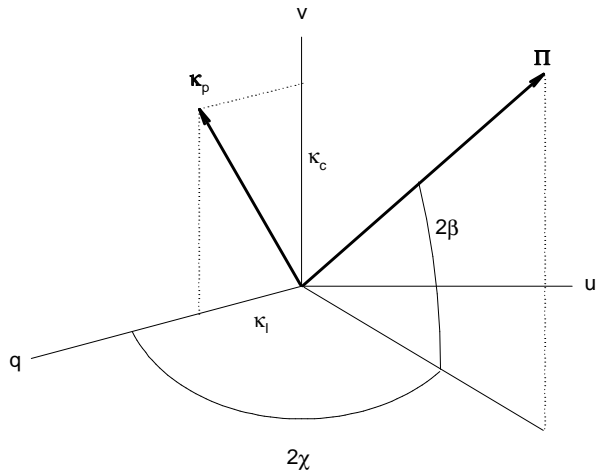,width=0.6\hsize,clip=}
\caption{The polarization vector \PI\ and the vector \kpvec\ that controls its
radiative transfer in the space defined by the normalized Stokes parameters $q
= Q/I$, $u = U/I$ and $v = V/I$.} \label{fig:q-u-v}
\end{figure}

The preceding discussion provides a geometric derivation of the structure of
polarization consistent with the fundamental physical processes that generate
maser radiation. Equation \ref{eq:GKK} lists the only polarization consistent
with the constraints that govern an interacting mixture of quantized particles
and classical electromagnetic waves that have equal amplitudes in the
quantization frame. In actuality, we cannot know these amplitudes beforehand
and must derive them from a complete solution for the level populations coupled
to the polarized radiation. The radiative transfer equation involves a matrix
in the space of four Stokes parameter, presenting a rather complex problem. An
elegant geometrical interpretation was derived by Litvak (1975). The
polarization structure of any electromagnetic wave is defined by the 3-vector
of its normalized Stokes parameters $\PI = (q,u,v)$ (figure \ref{fig:q-u-v}).
The off-diagonal elements of the radiative transfer matrix are $\kl \propto
\half(n_1^+ + n_1^-) - n_1^0$ and $\kc \propto n_1^+ - n_1^-$, where $n_1^m$ is
the population of the magnetic $m$-state of the upper level. These elements can
be combined to form another 3-vector $\kpvec = (\kl, 0, \kc)$, then
\eq{\label{eq:rotation}
    {d\PI\over d\ell} = \left[\PI\x\kpvec\right]\x\PI\,.
}
The effect of radiative transfer is to rotate the polarization vector of each
individual electromagnetic wave at the rotation velocity $\PI\x\kpvec$; this
velocity is different for different waves and changes as \PI\ is rotating.
Since waves are launched with arbitrary initial polarizations that subsequently
rotate at different rates, the radiation field can be expected to remain
unpolarized unless there is a stationary configuration whose polarization
vector does not rotate. When such a configuration exists, the polarization
vectors are locked once they enter it and that becomes the polarization of the
overall radiation field.

Stationary polarization obviously occurs when $\PI\ \|\ \kpvec$\,. However,
since the level populations are affected by the interaction with the maser
radiation, \kpvec\ itself varies and is affected by $\PI$, therefore one must
find a formalism to identify the stationary polarizations.  It is
straightforward to show that they are the eigenvectors of the radiative
transfer matrix (Elitzur 1996).  Two types of solutions exist for masers in a
magnetic field. One type corresponds to the $\xb \gg 1$ case, the other to $\xb
\ll 1$. The solution for the latter reproduces the GKK linear polarization
(\ref{eq:GKK}) accompanied by the circular polarization
\eq{\label{eq:v}
        v = {16x\xb\over3\cos\theta}\,.
}
This polarization arises when the populations of the magnetic sub-states of
each level become equal to each other as a result of maser radiative
interaction, requiring a unique cooperation between the particles and the
radiation that they amplify. It is reached only in masers with $\Js \gg S$,
where $S$ is the source function, after the radiation has grown so that $J/\Js
\ga \xb$. The condition $q^2 + v^2 \le 1$ constrains the propagation directions
for polarized maser radiation.

When the transition frequency varies, the Doppler width \DnuD\ varies
proportionately while the Zeeman splitting \nuB\ is unaffected.  Therefore \xb\
is inversely proportional to frequency, and the circular polarization decreases
with the transition frequency when all other properties remain fixed. McIntosh,
Predmore \& Patel (1994) find that SiO circular polarization indeed decreases
when the rotation quantum number, and with it transition frequency, increases.
The linear polarization displays the opposite trend, increasing with rotation
quantum number (McIntosh \& Predmore 1993). Since the solution linear
polarization is independent of transition wavelength, this is the expected
behavior in the presence of Faraday depolarization, which is proportional to
$\lambda^2$.  The low rotation states are more severely affected because of
their longer wavelengths and the linear polarization can be expected to
decrease toward lower angular momenta, as observed. Although detailed
calculations of Faraday rotation have yet to be performed for $\xb \ll 1$,
McIntosh \& Predmore find this to be the most plausible explanation of the
data.

\subsection{Limitations and Outstanding Issues}

The theory presented here was developed for an idealized maser. The results
depend in a crucial manner on the assumption of a constant direction for the
quantization axis. They provide the maximal polarization that can be produced
in a source that maintains a uniform direction for the magnetic field. Any
curvature in the field lines along the propagation direction results in
$\theta$ variations that destroy the phase coherence between the $\pi$- and
$\sigma$-components, reducing the degree of polarization. In particular, Alfven
waves introduce ripples in the field lines that destroy the polarization
whenever the Alfven wavelength is shorter than the amplification length.
Similarly, velocity blending will reduce circular polarization (Sarma, Troland
\& Romney, 2001) and linear polarization may also be reduced by Faraday
rotation. As a result, the information that can be extracted from polarization
alone is limited because the same polarization can be produced in a number of
different ways. For example, a certain linear polarization can be attributed
either to the maximal polarization at an appropriate angle $\theta$ or to a
higher degree of polarization that was degraded either by curvature in the
field lines or by Faraday depolarization.

Another fundamental assumption is that the only degeneracy of the maser levels
involves their magnetic sub-states. When any of the levels includes additional
degeneracy, so that magnetic sub-states of different levels overlap, the tight
constraints responsible for the stationary solutions no longer apply and the
polarization can be expected to disappear\footnote{By example, consider the
imaginary limit in which the hyperfine splitting of the OH molecule vanishes
and the four ground-state lines are blended into one.}. Indeed, the energy
levels of both \H2O and methanol involve hyperfine degeneracy and both masers
are generally only weakly polarized. Exceptions do exist, though, and \H2O
masers sometime display high polarization, notably during outbursts such as in
Orion (e.g., Abraham \& Vilas Boas 1994). This may involve the excitation of a
single hyperfine component, in which case the general solutions derived here
are applicable.

\paragraph{Incompleteness}
The theory presented here is incomplete. The radiation field is an ensemble of
waves launched with random polarizations. Subsequent maser amplification
through particle interactions is accompanied by rotation of each polarization
vector, and we have identified the stationary modes that do not rotate.
However, we have not shown how the radiation field actually evolves into this
solution, and this is considerably more difficult. Indeed, it is always simpler
to identify the stationary limit of a statistical distribution than to
demonstrate how this limit is actually attained. Demonstrations of the approach
to Maxwellian of a particle velocity distribution or to Planckian of a photon
distribution are considerably more difficult than the derivation of either
functional form. The evolution of such ensembles requires numerical simulations
of the type frequently performed in studies of plasma and laboratory lasers.
This approach can be avoided in the analysis of thermal radiation polarization,
where phases are meaningless. But the essence of the maser polarization
solution is specific phase relations among waves generated in different \Dm\
transitions, and those cannot be captured by standard radiative transfer
techniques---the very derivation of the radiative transfer equation from
Maxwell's equations is predicated on the assumption of random phases for
different waves (cf Litvak 1970, GKK). A full simulation of the ensemble
evolution of interacting particles and waves {\em is the only way to study
maser polarization growth}. Such simulations have not yet been attempted for
astronomical maser radiation. In addition to their inherent significance for
demonstrating the approach to stationary polarization, these simulations are
essential for a complete analysis of Faraday depolarization when $\nuB \ll
\DnuD$.

\vfil \acknowledgements

The support of an IAU travel grant and NSF grant AST-9819368 is gratefully
acknowledged.

\end{document}